\documentclass[letterpaper,10pt,doublecolumn]{IEEEtran}
\usepackage{xcolor}
\usepackage{multirow}
\usepackage{graphicx}
\usepackage{amsmath}
\usepackage{amssymb}
\usepackage{amsthm}
\usepackage{multirow}
\usepackage{diagbox}
\usepackage{colortbl}
\usepackage{booktabs}
\usepackage{arydshln}
\usepackage[colorlinks=true,linkcolor=black]{hyperref}

\usepackage{float}

\usepackage{subcaption}

\DeclareUnicodeCharacter{00A0}{ }

\begin{document}

\title{CLNet: Complex Input Lightweight Neural Network designed for Massive MIMO CSI Feedback}

\author{
Sijie~Ji,
Mo~Li,~\IEEEmembership{Fellow,~IEEE}

 \IEEEcompsocitemizethanks{
 This work is supported by Singapore MOE Tier 2 under grant T2EP20220-0011.
Sijie Ji and Mo Li are with School of Computer Science and Engineering, Nanyang Technological University, Singapore. (Emails: sijie001@e.ntu.edu.sg, limo@ntu.edu.sg).
}

}

\maketitle

\begin{abstract}
Unleashing the full potential of massive MIMO in FDD mode by reducing the overhead of CSI feedback has recently garnered attention. Numerous deep learning based massive MIMO CSI feedback approaches have demonstrate their efficiency and potential. However, most existing methods improve accuracy at the cost of computational complexity and the accuracy decreases significantly as the CSI compression rate increases.
This paper presents a novel neural network CLNet tailored for CSI feedback problem based on the intrinsic properties of CSI. CLNet proposes a forge complex-valued input layer to process signals and utilizes attention mechanism to enhance the performance of the network. The experiment result shows that CLNet outperforms the state-of-the-art method by average accuracy improvement of 5.41\% in both outdoor and indoor scenarios with average 24.1\% less computational overhead. Codes are available at GitHub. \footnote{\url{https://github.com/SIJIEJI/CLNet}}.

\end{abstract}


\IEEEpeerreviewmaketitle

\section{Introduction}

\IEEEPARstart{T}he massive multiple-input multiple-output (MIMO) technology is considered one of the core technologies of the next generation communication system, e.g., 5G. By equipping a large number of antennas, the base station (BS) can sufficiently utilize spatial diversity to improve the channel capacity. 
Especially, by enabling beamforming, a 5G BS can concentrate signal energy to a specific user equipment (UE) to achieve higher signal-to-noise ratio (SNR), less interference leakage and hence, higher channel capacity. 
However, beamforming is possibly conducted by the BS only when it has the channel state information (CSI) of the downlink at hand~\cite{marzetta2010noncooperative}. 

In frequency division duplexing (FDD) mode that most of contemporary cellular systems operate in, the channel reciprocity does not exist. Therefore, the UE would have to explicitly feed back the downlink CSI to the BS, and the pilot-aided training overhead grows quadratically with the number of transmitting antennas, which might overturn the benefit of Massive MIMO itself~\cite{lu2014overview}. Thus, CSI compression is needed before the feedback to reduce the overhead.

Traditional compressive sensing (CS) based methods rely heavily on channel sparsity and are limited by their efficiency in iteratively reconstructing the signals. Their performance is highly dependent on the wireless channel~\cite{kyritsi2003correlation}, and thus is not a desirable approach considering the diversified use cases of 5G networks.

The recent rapid development of deep learning (DL) technologies provide another possible solution for efficient CSI feedback in FDD massive MIMO system. Instead of relying on sparsity, the DL approaches utilize the auto-encoder framework~\cite{hinton2006reducing}. The encoder learns a map to the low-dimensional compressed space and the decoder reconstruct to the original data by a single run without the labeled data. It naturally overcomes the limits of CS-based approaches in channel sparsity and operation efficiency. 

The first DL-based method, CsiNet~\cite{wen2018deep}, explored and demonstrated the efficiency of deep learning in CSI feedback. CsiNet significantly outperforms the traditional CS-based methods (LASSO, BM3D-AMP and TVAL3) under various compression rates. 

Based on CsiNet, most of the subsequent DL-based methods utilize more powerful DL building blocks to achieve better performance with the sacrifice of computational overhead. CsiNet-LSTM~\cite{wang2018deep} and Attention-CSI~\cite{cai2019attention} introduced LSTM that significantly increases the computational overhead. CsiNet+~\cite{guo2020convolutional} comprehensively surveyed recent DL-based methods and proposed a parallel multiple-rate compression framework. The computational overhead of CsiNet+ are approximately x7 higher than the original CsiNet~\cite{lu2021binarized}. Recently, some methods start to reduce the complexity, for example, JCNet~\cite{lu2019bit} and BcsiNet~\cite{lu2021binary}, however, their performance has also reduced. So far, only CRNet~\cite{lu2020multi} has outperformed CsiNet without increasing the computational complexity.

However, CSI or signals are represented in complex envelopes, which have their own physical meaning that is overlooked by previous works, only~\cite{zhang2020cv} considered this problem by adopting complex-valued three dimensional convolutional neural network~\cite{dcomplex}. However, as the complex kernel is hard to optimize through back-propagation, the network is hard to train and the computational 
complexity is inevitably greatly increased. Considering the limited computational resource and limited storage at UE side, this letter proposes a tailored DL network that can cope with \textbf{c}omplex number yet maintain \textbf{l}ightweight, CLNet, for CSI feedback problem. Eventually, CLNet outperforms CRNet with 5.41\% higher accuracy and 24.1\% less complexity on average. The main contributions are summarized as follows:

\begin{itemize}
\item CLNet proposes a simple yet effective way to organic integrate the real and imaginary parts into the real-valued neural network models.
\item CLNet adopts spatial attention mechanisms to let the DL model focus on the more informative clustered signal parts.

\end{itemize}

\section{System Model and Preliminary}

Consider a single cell FDD system using massive MIMO with $N_{t}$ antennas at BS, where $N_{t}$ $\gg 1$ and $N_{r}$ antennas at UE side ($N_{r}$ equals to 1 for simplicity). The received signal $\mathit{y} \in \mathbb{C}^{ N_{c} \times 1}$ can be expressed as
\begin{equation}
\mathit{y}=\textit{\textbf{A}} \mathit{x}+\mathit{z}
\label{e:received_signal}
\end{equation}
where $N_{c}$ indicates the number of subcarriers, $\mathit{x} \in \mathbb{C}^{N_{c} \times 1}$ indicates the transmitted symbols,  and $\mathit{z} \in \mathbb{C}^{N_{c} \times 1}$ is the complex additive Gaussian noise. 
$\textbf{A}$ can be expressed as $\operatorname{diag}\left(\mathit{h}_{1}^{H} \mathit{p}_{1}, \cdots, \mathit{h}_{N_{c}}^{H} \mathit{p}_{N_{c}}\right)$, where $\mathit{h}_{i} \in \mathbb{C}^{N_{t} \times 1}$ and $\mathit{p}_{i} \in \mathbb{C}^{N_{t} \times 1}, i \in\left\{1, \cdots, N_{c}\right\}$ represent downlink channel coefficients and beamforming precoding vector for subcarrier $i$, respectively.

In order to derive the beamforming precoding vector $\mathit{p}_{i}$, the BS needs the knowledge of corresponding channel coefficient $\mathit{h}_{i}$, which is fed back by the UE. Suppose that the downlink channel matrix is $\textit{\textbf{H}}=\left[\mathit{h}_{1} \cdots \mathit{h}_{N_{c}}\right]^{H}$ which contains $N_{c}N_{t}$ elements. The number of parameters that need to be fed back is $2N_{c}N_{t}$, including the real and imaginary parts of the CSI, which is proportional to the number of antennas.

Because the channel matrix $\textit{\textbf{H}}$ is often sparse in the angular-delay domain. By 2D discrete Fourier transform (DFT), the original form of spatial-frequency domain CSI can be converted into angular-delay domain, such that
\begin{equation}
\textit{\textbf{H}}^{\prime}=\textit{\textbf{F}}_{c} \textit{\textbf{H}} \textit{\textbf{F}}_{t}^{H}
\end{equation}
where $\textit{\textbf{F}}_{c}$ and $\textit{\textbf{F}}_{t}$ are the DFT matrices with dimension $N_{c} \times N_{c}$ and $N_{t} \times N_{t}$, respectively. For the angular-delay domain channel matrix $\textit{\textbf{H}}^{\prime}$, every element corresponds to a certain path delay with a certain angle of arrival (AoA). In $\textit{\textbf{H}}^{\prime}$, only the first $N_{a}$ rows contain useful information, while the rest rows represent the paths with larger propagation delays are made up of near-zero values, can be omitted without much information loss. 
Let $\textit{\textbf{H}}_{a}$ denote the informative rows of $\textit{\textbf{H}}^{\prime}$.

$\textit{\textbf{H}}_{a}$ is input into UE's encoder to produce the codeword $\textit{\textbf{v}}$ according to a given compression ratio $\eta$ such that
\begin{equation}
\textit{\textbf{v}}=f_\mathcal{E}\left(\textit{\textbf{H}}_{a}, \mathit{\Theta}_{\mathcal{E}}\right)
\end{equation}
where $f_{\mathcal{E}}$ denotes the encoding process and $\mathit{\Theta}_{\mathcal{E}}$ represents a set of parameters of the encoder.

Once the BS receives the codeword $\textit{\textbf{v}}$, the decoder is used to reconstruct the channel by 
\begin{equation}
\hat{\textit{\textbf{H}}}_{a}=f_{\mathcal{D}}\left(\textit{\textbf{v}}, \mathit{\Theta}_{\mathcal{D}}\right)
\end{equation}
where $f_{\mathcal{D}}$ denotes the decoding process and $\mathit{\Theta}_{\mathcal{D}}$ represents a set of parameters of the decoder. 
Therefore, the entire feedback process can be expressed as
\begin{equation}
\hat{\textit{\textbf{H}}}_{a}=f_\mathcal{D}\left(f_\mathcal{E}\left(\textit{\textbf{H}}_{a}, \mathit{\Theta}_{\mathcal{E}}\right), \mathit{\Theta}_{\mathcal{D}}\right)
\end{equation}
The goal of CLNet is to minimize the difference between the original $\textit{\textbf{H}}_{a}$ and the reconstructed $\hat{\textit{\textbf{H}}}_{a}$, which can be expressed formally as finding the parameter sets of encoder and decoder satisfying
\begin{equation}
\left(\hat{\mathit{\Theta}}_{\mathcal{E}}, \hat{\mathit{\Theta}}_{\mathcal{D}}\right)=\underset{\mathit{\Theta}_{\mathcal{E}}, \mathit{\Theta}_{\mathcal{D}}}{\arg \min }\left\|\textit{\textbf{H}}_{a}-f_{\mathcal{D}}\left(f_{\mathcal{E}}\left(\textit{\textbf{H}}_{a}, \mathit{\Theta}_{\mathcal{E}}\right), \mathit{\Theta}_{\mathcal{D}}\right)\right\|_{2}^{2}
\end{equation}

\section{CLNet Design}

\begin{figure}[h]
  \centering
  \includegraphics[width=\linewidth]{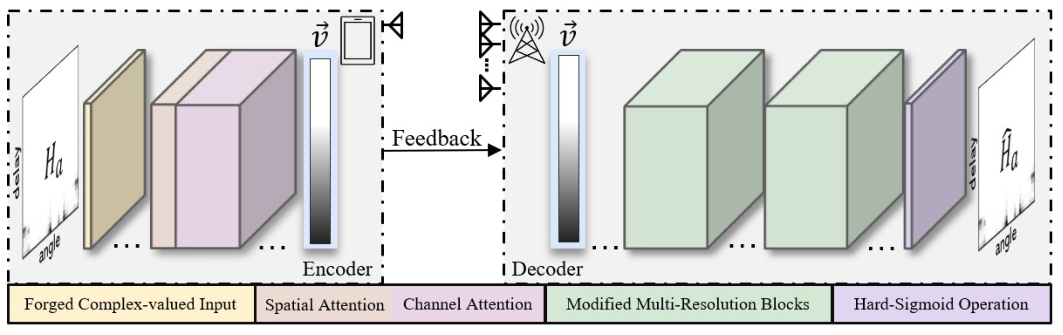}
  \caption{The encoder and decoder architecture of CLNet.}
  \label{f:encoder}
\end{figure}

This section presents the design of the CLNet and its key components. Figure~\ref{f:encoder} depicts the overall architecture of CLNet, in which traditional convolution blocks are omitted for simplicity. Overall, CLNet is an encoder-decoder framework with four main building blocks that are tailored for the CSI feedback problem. 

The performance of the CSI feedback scheme highly depends on the compression part, the encoder. The less information loss of the compression, the higher the decompression accuracy can be obtained. Due to the limited computing power and storage space of UE, deepening the encoder network design is not practical. Therefore, CLNet leverages the physical characteristics of CSI to achieve a lightweight yet informative encoder by two tailored blocks. 

First, CSI is the channel frequency response with complex values that depict channel coefficients of different signal paths. The previous DL-based CSI feedback methods, treat the real and imaginary parts of the CSI separately. Instead, the input CSI in CLNet first goes through the forged complex-valued input layer that embeds the real and imaginary parts together to preserve the physical information of the CSI (Section III-A). Second, different signal paths have different resolutions of cluster effect in the angular-delay domain, which corresponding to different angles of arrival and different path delays. Thus, we introduce the CBAM block~\cite{woo2018cbam} that serves as spatial-wise attention to force the neural network focus on those clusters and suppress the unnecessary parts (Section III-B).

Since the encoder becomes more powerful, the decoder can be correspondingly more lightweight, thus CLNet modifies the CRBlocks~\cite{lu2020multi} in decoder by reducing the filter size from $1\times9$ to $1\times3$. To further reduce the computational cost, CLNet adopts the hard-Sigmoid activation fuction which is more hardware friendly than the conventional Sigmoid activation function (Section III-C). 

\subsection{Forged Complex-valued Input}

CSI is complex-valued channel coefficients such that:
\begin{equation}
\textit{\textbf{H}}(t)=\sum_{k=1}^{N} a_{k}(t) e^{-j\theta_{k}(t)}
\end{equation}
where $N$ is the number of signal paths. $a_{k}(t)$ and $\theta_{k}(t)$ indicate the signal attenuation and propagation phase rotation of the $k$-th path at time $t$ respectively. The BS relies the physical meaning of CSI, the norm of real and imaginary part describes the channel's attenuation to signal and the ratio of the real and the imaginary part describes the channel's phase rotation to the signal, to conduct the beamforming.

\begin{figure}[h]
  \centering
  \includegraphics[width=\linewidth,height=2.9cm]{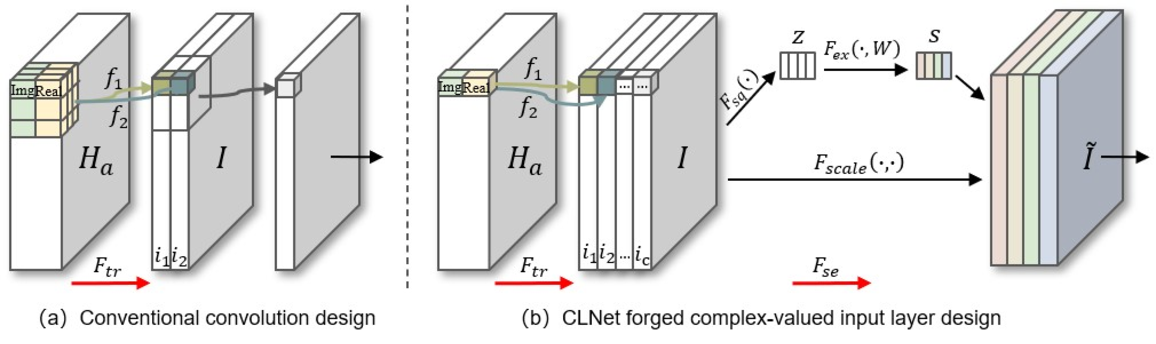}
  \caption{Diagrammatic comparison of the conventional convolution and the CLNet forged complex-valued input layer.}

  \label{f:input}
\end{figure}

Since a typical deep learning neural network is designed based on real-valued inputs, operations, and representations.  Existing DL-based CSI feedback methods simply separate the real and imaginary parts of the complex-valued as two independent channels of an image as the neural network input, which may destroy the original physical property of each complex-valued channel coefficient. 
Specifically,  as Figure~\ref{f:input} (a) depicts, a conventional $3 \times 3$ kernel size entangles the real and imaginary parts of neighboring elements in $\textit{\textbf{H}}_{a}$, and as a result, the 9 complex CSI are interpolated as one synthesized value. 

Mathematically, $\textit{\textbf{F}}_{t r}: \textit{\textbf{H}}_{a} \rightarrow \mathcal{I}$ is a convolutional transformation. Here, $\textit{\textbf{H}}_{a} \in \mathbb{R}^{N_a \times N_a \times 2}$ is a 3D tensor, extended from its 2D version by including an additional dimension to separately express the real and imaginary parts, and $\mathcal{I} \in \mathbb{R}^{N_a \times N_a \times C}$, where $C$ indicates the number of convolutional filters applied to learn different weighted representations. The output of $\textit{\textbf{F}}_{t r}$ is $\mathcal{I}=\left[\textit{\textbf{i}}_{1}, \textit{\textbf{i}}_{2}, \ldots, \textit{\textbf{i}}_{C}\right]$, $\textit{\textbf{i}}_{c} \in  \mathbb{R}^{N_a \times N_a}$.  Let $a_{n}+b_{n}i$ denotes a CSI and $w_{n}$ is the learnable weight of a convolutional filter $f$. The 3x3 convolution operation essentially is the sum of two multiplication such that:
\begin{equation}
   i_{1}(1,1) = [a_{1},...,a_{9}] \cdot [w_{1},...,w_{9}] + [b_{1},...,b_{9}] \cdot [w_{1},...,w_{9}] 
\end{equation}
In such way, the real and imaginary parts of the same complex-valued signal are decoupled and different CSI metrics are mixed, thus losing the original physical information carried by the channel matrix.

The insight of CLNet is that by utilizing a $1\times1$ point-wise convolution, the real and imaginary parts of a complex-valued coefficient can be explicitly embedded such that:
\begin{equation}
   i_{1}(1,1) = [a_{1}] \cdot [w_{1}] + [b_{1}] \cdot [w_{1}] 
\end{equation}
where the ratio between $a$ and $b$ are preserved, thus maintain the phase information and the amplitude of the signal be scaled by $w$. Since CNN shares the weight $w$, so the entire whole CSI matrix’s amplitude is essentially scaled by the same $w$, the relative amplitude across subchannels is also preserved.

The output $i_{c}$, essentially, is a weighted representation of the original $\textit{\textbf{H}}_{a}$ and different filters learn different weighted representations, among which, some may be more important than others. Based on this, CLNet further adopts the SE block~\cite{hu2018squeeze}, which serve as the channel-wise attention in the forged complex-valued input layer. It assists the neural network to model the relationship of the weights so as to focus on the important features and suppress the unnecessary ones. A diagram of the SE block is shown in Figure~\ref{f:input} (b) with annotation $\textit{\textbf{F}}_{se}$. 

The output $\mathcal{I}$ first goes through $\textit{\textbf{F}}_{s q}$ transformation by global average pooling to obtain channel-wise statistics descriptor $\textit{\textbf{z}} \in \mathbb{R}^{C}$. Here, $\textit{\textbf{F}}_{s q}$ expands the neural network receptive field to the whole angular-delay domain to obtain the global statistical information, compensating the shortcoming of the insufficient local receptive field of $1\times1$ convolution used in the first step of the forged complex-valued input layer.

After that, the channel descriptor $\textit{\textbf{z}}$ goes through $\textit{\textbf{F}}_{e x}$ transformation, i.e., a gated layer with sigmoid activation to learn the nonlinear interaction as well as the non-mutually-exclusive relationship between channels, such that
\begin{equation}
\mathbf{s}=\textit{\textbf{F}}_{e x}(\textit{\textbf{z}}, \textit{\textbf{W}})=\sigma(g(\textit{\textbf{z}}, \textit{\textbf{W}}))=\sigma\left(\textit{\textbf{W}}_{2} \delta\left(\textit{\textbf{W}}_{1} \textit{\textbf{z}}\right)\right)
,
\end{equation}
where $\delta$ is the ReLU function, $\textit{\textbf{W}}_{1} \in \mathbb{R}^{\frac{C}{2} \times C}$ and $\textit{\textbf{W}}_{2} \in \mathbb{R}^{C \times \frac{C}{2}}$. $\textit{\textbf{F}}_{e x}$ further explicitly models the inter-channel dependencies based on $\textit{\textbf{z}}$ and obtain the calibrated $\mathbf{s}$, which is the attention vector that summarizes all the characteristics of channel $C$, including intra-channel and inter-channel dependencies. Before being fed into the next layer, each channel of $\mathcal{I}$ is scaled by the corresponding attention value, such that
\begin{equation}
\tilde{\mathcal{I}}_{:,:,i}=\textit{\textbf{F}}_{scale}(\mathbf{s},\mathcal{I})=\mathbf{s}_{i} \mathcal{I}_{:,:,i}, \text { s.t. } i \in\{1,2, \cdots, C\}
\end{equation}
$\tilde{\mathcal{I}} \in \mathbb{R}^{N_a \times N_a \times C}$ is the final output of the forged complex-valued input layer, which preserves the CSI physical information while capturing dynamics by the channel-wise attention mechanism.

\subsection{Attention Mechanism for Informative Encoder}
\begin{figure}[h]
    \centering
    \includegraphics[width=\linewidth]{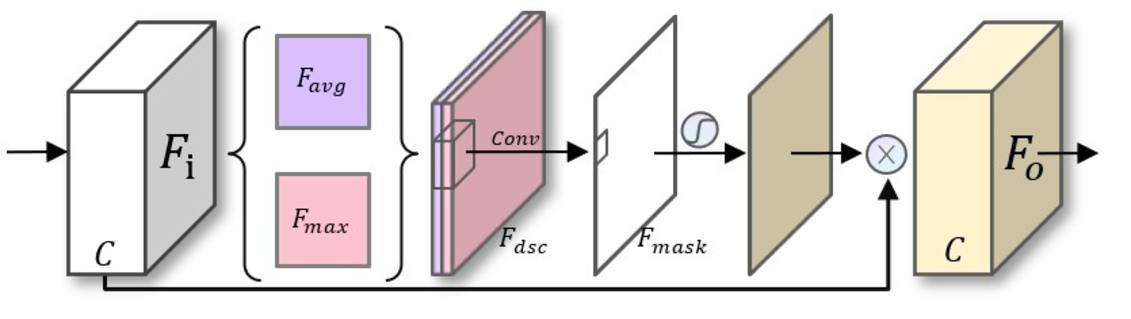}
    \caption{Operation illustration of spatial-wise attention of CLNet.}
  \label{f:cbam}
\end{figure}

On the other hand, in angular-delay domain, the channel coefficients exhibit the effect of clusters with different resolutions that corresponding to the distinguishable paths that arrive with specific delays and AoAs. In order to pay more attention to such clusters, CLNet employs a CBAM block~\cite{woo2018cbam} serve as spatial-wise attention to distinguish them with weights in the spatial domain as Figure~\ref{f:cbam} illustrates. 

Based on the cluster effect in the angular-delay domain, spatial-wise attention uses the generated spatial statistical descriptors as the basis for assigning weights, forcing the network to focus more on the distinguishable propagation paths.

First, two pooling operations, i.e., average-pooling and max-pooling, are adopted across the input $\textit{\textbf{F}}_{i}$'s channel $C$ to generate two 2D feature maps,  $\textit{\textbf{F}}_{\mathbf{a v g}} \in \mathbb{R}^{ N_{a} \times N_{a} \times 1}$ and
$\textit{\textbf{F}}_\mathbf{m a x} \in \mathbb{R}^{ N_{a} \times N_{a}\times 1}$, respectively.
CLNet concatenates the two feature maps to generate a compressed spatial feature descriptor $\textit{\textbf{F}}_\mathbf{d s c} \in \mathbb{R}^{ N_{a} \times N_{a}\times 2}$, and convolves it with a standard convolution layer to produce a 2D spatial attention mask $\textit{\textbf{F}}_\mathbf{m a s k} \in \mathbb{R}^{ N_{a} \times N_{a}\times 1}$. The mask is activated by Sigmoid and then multiplied with the original feature maps $\textit{\textbf{F}}_{i}$ to obtain $\textit{\textbf{F}}_{o}$ with spatial-wise attention. 
\begin{equation}
\begin{aligned}
\textit{\textbf{F}}_{o} &= \mathbf{CBAM}(\textit{\textbf{F}}_{i})\\ %
&=\textit{\textbf{F}}_{i}\left(\sigma\left(\textit{\textbf{f}}_{c}([\operatorname{AvgPool}(\textit{\textbf{F}}_{i}) ; \operatorname{Max} \operatorname{Pool}(\textit{\textbf{F}}_{i})])\right)\right) \\
&=\textit{\textbf{F}}_{i}\left(\sigma\left(\textit{\textbf{f}}_{c}\left(\left[\textit{\textbf{F}}_{\mathbf{a v g}} ; \textit{\textbf{F}}_{\mathbf{m a x} }\right]\right)\right)\right)
\end{aligned}
\end{equation}
With spatial-wise attention, CLNet focuses the neural network on the more informative signal propagation paths in the angular-delay domain.

\subsection{Reduction of Computational Cost}

The often-used Sigmoid activation function contains exponential operation:
\begin{equation}
\sigma(x)=\frac{1}{1+e^{-x}}=\frac{e^{x}}{e^{x}+1}.
\end{equation}
%
In order to reduce time cost in the computation, CLNet uses the hard version of Sigmoid, its piece-wise linear analogy function, denoted as $h\sigma$ to replace the Sigmoid function~\cite{howard2019searching},
\begin{equation}
h\sigma(x)=\frac{\min (\max (x+3, 0), 6)}{6}
\end{equation}

\begin{figure}[h]
  \centering
  \includegraphics[scale=0.3]{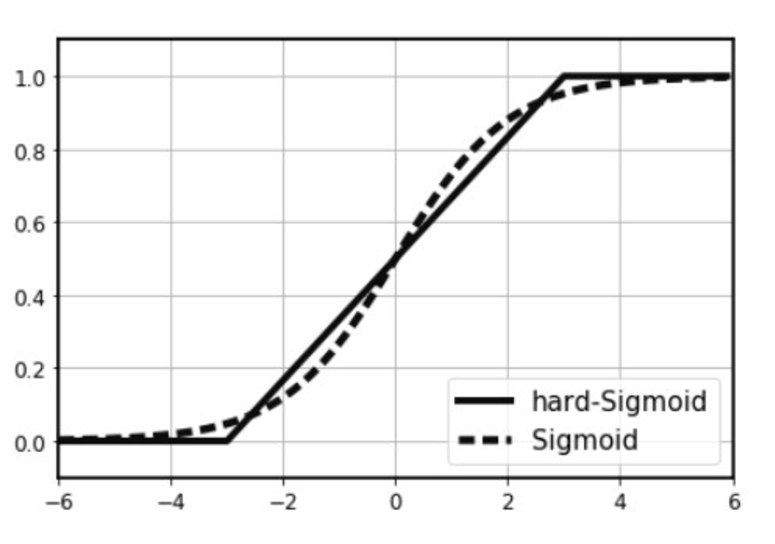}
  \caption{Comparison between Sigmoid and hard-Sigmoid functions.}
  \label{f:hsigg}
\end{figure}

Figure~\ref{f:hsigg} compares the excitation curves of the hard-Sigmoid and Sigmoid functions.
The hard-Sigmoid induces no discernible degradation in the accuracy but benefits from its computational advantage of entailing no exponential calculations. 
In practice, the hard-Sigmoid can fit in most software and hardware frameworks and can mitigate the potential numerical quantization loss introduced by different hardware.

\section{Evaluation}

This section presents the detailed experiment setting and the comparison with the state-of-the-art (SOTA) DL-based CSI feedback approach, in terms of accuracy and computational overhead.
\subsubsection{\textbf{Data Generation}} 

To ensure a fair performance comparison, we use the same dataset as provided in the first work of DL-based Massive MIMO CSI feedback in~\cite{wen2018deep}, which is also used in later studies on this problem~\cite{wang2018deep,lu2021binarized,guo2020convolutional,lu2020multi,cai2019attention}.
The generated CSI matrices are converted to angular-delay domain $\textit{\textbf{H}}_{a} \in \mathbb{R}^{32 \times 32 \times 2}$ by 2D-DFT. 
The total 150,000 independently generated CSI are split into three parts, i.e., 100,000 for training, 30,000 for validation, and 20,000 for testing, respectively.

\subsubsection{\textbf{Training Scheme and Evaluation Metric}}

The normalized mean square error ($\mathrm{NMSE}$) between the original $\textit{\textbf{H}}_{a}$ and the reconstructed $\hat{ \textit{\textbf{H}}}_{a} $ is used to evaluate the network accuracy:
\begin{equation}
    \mathrm{NMSE}=\mathrm{E}\left\{\|\textit{\textbf{H}}_{a}-\hat{\textit{\textbf{H}}}_{a}\|_{2}^{2} /\|\textit{\textbf{H}}_{a}\|_{2}^{2}\right\}
\end{equation}
The complexity is measure by the flops (floating-point operations per second).
%
The model was trained with the batch size of 200 and 8 workers on a single NVIDIA 2080Ti GPU. The epoch is set to 1000, as recommended in previous work~\cite{lu2020multi,guo2020convolutional}. To further ensure the fairness, we fix the random seed of the computer.
\subsubsection{\textbf{CLNet Overall Performance}}

\begin{table*}
\small\addtolength{\tabcolsep}{-1.5pt}
\centering
\caption[Caption for LOF]
{NMSE(dB)$^a$ and complexity comparison between series of CSI feedback network and the proposed CLNet.}
\footnotesize{$^a$ / means the performance is not reported.}
\begin{tabular}{l|rrr|rrr|rrr|rrr|rrr} 
\hline\hline
\multicolumn{1}{c}{$\eta$}                      & \multicolumn{3}{c}{1/4}                                                                          & \multicolumn{3}{c}{1/8}                                                                       & \multicolumn{3}{c}{1/16}                                                                      & \multicolumn{3}{c}{1/32}                                                                      & \multicolumn{3}{c}{1/64}                                                                        \\ 
\hline\hline
\multicolumn{1}{c}{\multirow{2}{*}{Methods}} & \multicolumn{1}{c}{\multirow{2}{*}{FLOPS}} & \multicolumn{2}{c}{NMSE}                            & \multicolumn{1}{c}{\multirow{2}{*}{FLOPS}} & \multicolumn{2}{c}{NMSE}                         & \multicolumn{1}{c}{\multirow{2}{*}{FLOPS}} & \multicolumn{2}{c}{NMSE}                         & \multicolumn{1}{c}{\multirow{2}{*}{FLOPS}} & \multicolumn{2}{c}{NMSE}                         & \multicolumn{1}{c}{\multirow{2}{*}{FLOPS}} & \multicolumn{2}{c}{NMSE}                           \\
\multicolumn{1}{c}{}                         & \multicolumn{1}{c}{}                       & \multicolumn{1}{c}{indoor}   & \multicolumn{1}{c}{outdoor}  & \multicolumn{1}{c}{}                       & \multicolumn{1}{c}{indoor} & \multicolumn{1}{c}{outdoor} & \multicolumn{1}{c}{}                       & \multicolumn{1}{c}{indoor} & \multicolumn{1}{c}{outdoor} & \multicolumn{1}{c}{}                       & \multicolumn{1}{c}{indoor} & \multicolumn{1}{c}{outdoor} & \multicolumn{1}{c}{}                       & \multicolumn{1}{c}{indoor}  & \multicolumn{1}{c}{outdoor}  \\ 
\hline\hline
CLNet                                        & \textbf{\textit{4.05M}}                   & \textbf{\textit{-29.16}} & \textbf{\textit{-12.88}} & \textbf{\textit{3.01M}}                    & \textbf{-15.60}        & \textbf{-8.29}          & \textbf{\textit{2.48M}}                    & \textbf{-11.15}        & \textbf{-5.56}          & \textbf{\textit{2.22M}}                    & \textbf{-8.95}         & \textbf{-3.49}          & \textbf{\textit{2.09M}}                    & \textbf{\textit{-6.34}} & \textbf{-2.19}           \\
CRNet                                        & 5.12M                                      & -24.10                   & -12.57                   & 4.07M                                      & -15.04                 & -7.94                   & 3.55M                                      & -10.52                 & -5.36                   & 3.29M                                      & -8.90                  & -3.16                   & 3.16M                                      & -6.23                   & \textbf{-2.19}           \\
CSINet\cite{wen2018deep}                                       & 5.41M                                      & -17.36                   & -8.75                    & 4.37M                                      & -12.70                 & -7.61                   & 3.84M                                      & -8.65                  & -4.51                   & 3.58M                                      & -6.24                  & -2.81                   & 3.45M                                      & -5.84                   & -1.93                    \\ 
\hdashline
CSINet+\cite{guo2020convolutional}                                      & 24.57M                                     & -27.37                   & -12.40                   & 23.52M                                     & \textit{-18.29}        & \textit{-8.72}          & 23.00M                                     & \textit{-14.14}        & -5.73                   & 22.74M                                     & \textit{-10.43}        & -3.40                   & 22.61M                                     & \multicolumn{1}{c}{/}   & \multicolumn{1}{c}{/}    \\
Attn-CSI\cite{cai2019attention}                                     & 24.72M                                     & -20.29                   & -10.43                   & 22.62M                                     & \multicolumn{1}{c}{/}  & \multicolumn{1}{c|}{/}  & 21.58M                                     & -10.16                 & \textit{-6.11}          & 21.05M                                     & -8.58                  & \textit{-4.57}          & 20.79M                                     & -6.32                   & \textit{-3.27}          
\end{tabular}
\label{t:all}

\end{table*}

Table~\ref{t:all} shows the overall performance comparison among the proposed CLNet and related CSI feedback networks.

As for the complexity, generally, the LSTM-based networks (CSINet+ and Attn-CSI) require approximate five to seven-flods higher computational resources than the CNN-based networks (CSINet, CRNet\footnote{Note that the CRNet paper reported flops is corrected by [13].} and CLNet). Furthermore, because LSTM's operation relies on the previous output as the input of the hidden layer and cannot share parameters for parallel computation, it is difficult to reduce the complexity even if the compression rate increases.
As we can see from Table~\ref{t:all}, the CLNet is the lightest among all these networks. Compared with the SOTA CRNet, the CLNet significantly reduces the computational complexity by 24.1\% fewer flops on average. The flops of CLNet is 18.00\%, 22.35\%, 25.20\%, 26.50\%, 28.36\% less than CRNet at the compression ratio $\eta$ of 1/64, 1/32, 1/16, 1/8, 1/4, respectively. As the compression rate increases, the computational complexity degrades more. 

Turn to the accuracy part, the best results in the lightweight network are shown in bold, and the best results in all networks are shown in italics. For the accuracy as shown in Table~\ref{t:all}, the result shows that CLNet consistently outperforms other lightweight networks at all compression ratios in both indoor and outdoor scenarios with 5.41\% overall average improvement compared with the SOTA CRNet\footnote{We reproduce CRNet follow the open source code: https://github.com/Kylin9511/CRNet the higher performance they reported in the paper are from training with 2500 epoch.}. In indoor scenarios, CLNet obtains an average performance increase of 6.61\%, with the most increase of 21.00\% at the compression ratio of $\eta=1/4$. In outdoor scenarios, the average improvement on NMSE is 4.21\%, with the most increase of 10.44\% at the compression ratio of $\eta=1/32$. 
Compared to heavyweight networks, CLNet still achieves the best results at the compression ratio of 1/4, outperforming the second place CSINet+ by 6.54\% and 3.87\% in indoor and outdoor scenario respectively. CLNet also achieves the best result in indoor scenario at the compression ratio equals to 1/64.
\subsubsection{\textbf{Ablation Study}}
Considering the limited interpretability of deep neural network, we further conduct the
ablation study to better quantify the gain of the proposed forged complex-valued input layer and spatial-attention mechanism. The epochs of ablation studies are set to 500 in indoor scenarios, the rest settings remain the same as discussed in \S IV(1-2). Baseline is the CRNet with conventional convolution. 

\begin{table}[h]
\small\addtolength{\tabcolsep}{-4.5pt}
\centering
\caption{NMSE (dB) Comparison of Ablation Study.}
\label{t:ab}
\begin{tabular}{r|r|r|r|r|r} 
\hline\hline
\multicolumn{1}{c}{$\mathbf{\eta}$} & \multicolumn{1}{c}{\textbf{Baseline }} & \multicolumn{1}{c}{\textbf{ 1x1 Conv}} & \multicolumn{1}{c}{\begin{tabular}[c]{@{}c@{}}\textbf{ 1x1 Conv}\\\textbf{+ SE}\end{tabular}} & \multicolumn{1}{c}{\begin{tabular}[c]{@{}c@{}}\textbf{1x1 Conv}\\\textbf{+ CBAM }\end{tabular}} & \multicolumn{1}{c}{\begin{tabular}[c]{@{}c@{}}\textbf{1x1 Conv}\\\textbf{+ SE + CBAM }\end{tabular}}  \\ 
\hline\hline
1/4                        & -21.702                                & -27.694                                & -27.903                                                                                       & -28.142                                                                                         & -28.984                                                                                               \\ 
\hline
1/8                        & -13.037                                & -15.171                                & -15.167                                                                                       & -15.321                                                                                         & -15.487                                                                                               \\ 
\hline
1/16                       & -10.212                                & -11.013                                & -11.231                                                                                       & -10.684                                                                                         & -11.217                                                                                               \\ 
\hline
1/32                       & -8.443                                 & -8.525                                 & -8.732                                                                                        & -8.613                                                                                          & -8.885                                                                                                \\ 
\hline
1/64                       & -6.023                                 & -6.145                                 & -6.201                                                                                        & -6.086                                                                                          & -6.297                                                                                                \\
\hline
\end{tabular}
\end{table}
As Table~\ref{t:ab} shown, by simply modifying the first layer from a conventional convolution layer to an 1x1 convolution as the forged complex input layer,its accuracy surpasses the baseline at all compression ratios with an average improvement of 10.964\%, which demonstrates the efficacy of appropriately preserving the complex notation. After adding the SE block, the accuracy is slightly improved although there is no improvement at $\eta=1/8$.
The last two columns show that the spatial-attention slightly improves the accuracy at low compression rates, however, when combined with the SE block, its accuracy is further improved by 3.058\% on average. 

\subsubsection{\textbf{Encoder Complexity}}
Table~\ref{t:enc} reveals that the CLNet encoder is actually slightly heavier than that of CRNet. However, the BS may need to execute several different models at the same time so a relatively light decoder would also be beneficial. In terms of storage space, CLNet and CRNet are roughly the same.

\begin{table}[h]
\centering
\caption{ Detailed Complexity of CRNet and CLNet}
\begin{tabular}{r|c|cc|cc} 
\hline\hline
\multicolumn{1}{c}{\multirow{2}{*}{$\mathbf{\eta}$}} & \multicolumn{1}{c}{\multirow{2}{*}{\textbf{Method}}} & \multicolumn{2}{c}{\textbf{Encoder at UE}}       & \multicolumn{2}{c}{\textbf{Decoder at BS}}  \\
\multicolumn{1}{c}{}                     & \multicolumn{1}{c}{}                        & flops(M) & \multicolumn{1}{c}{\#params} & flops(M) & \#params                \\ 
\hline\hline
\multirow{2}{*}{1/4}                     & CLNet                                       & 1.34     & 1.049M                       & 2.71     & 1.052M                  \\
                                         & CRNet                                       & 1.20     & 1.049M                       & 3.92     & 1.053M                  \\ 
\hline
\multirow{2}{*}{1/64}                    & CLNet                                       & 0.36     & 65.954K                      & 1.73     & 69.210K                 \\
                                         & CRNet                                       & 0.22     & 65.720K                      & 2.94     & 70.386K                 \\
\hline
\end{tabular}
\label{t:enc}
\end{table}
\section{Conclusion}

This article studies the CSI feedback problem for massive MIMO under FDD mode, which is the key technology of 5G communication systems. Based on the understanding of the physical properties of the CSI data, a novel customized deep learning framework, CLNet, is proposed. The forged complex-valued input layer preserves the amplitude and phase information of the signal and enhances with spatial-attention mechanisms. The hard-Sigmoid function is adopts to eliminate the exponential calculations.
The overall performance of CLNet has 5.41\% higher accuracy than the state-of-the-art CRNet with 24.10\% less computation overhead.

\bibliographystyle{IEEEtran}
\bibliography{bitex}

\end{document}